\documentclass[useAMS,usenatbib]{mn2e}

\usepackage[active]{srcltx}
\usepackage{graphicx}
\usepackage{txfonts}
\usepackage{natbib}
\usepackage{multirow}
\usepackage{longtable}
\usepackage[applemac]{inputenc}
\usepackage{rotating}

\newcommand{\civ} {\mbox{C$\;$\sc{iv}}}
\newcommand{\oii} {\mbox{O$\;$\sc{ii}}}

\newcommand{\msun}{\ensuremath{\mbox{M}_{\odot}}}
\newcommand{\rsun}{\ensuremath{\mbox{R}_{\odot}}}

\newcommand{\kms}{km.s$^{-1}\,$}

\def\gtrsim{\mathrel{\hbox{\rlap{\hbox{\lower4pt\hbox{$\sim$}}}\hbox{$>$}}}}
\def\ltsim{\mathrel{\hbox{\rlap{\hbox{\lower4pt\hbox{$\sim$}}}\hbox{$<$}}}}

\title[Magnetic oblique rotator model for HD~191612]{Confirmation of the magnetic oblique rotator model\\for the Of?p star HD~191612\thanks{Based on observations obtained at the Canada-France-Hawaii Telescope (CFHT) which is operated by the National Research Council of Canada, the Institut National des Sciences de l'Univers of the Centre National de la Recherche Scientifique of France, and the University of Hawaii}}
\author[G.A. Wade et al.]{G.A. Wade$^{1}$, I.D. Howarth$^{2}$, R.H.D. Townsend$^{3}$, J.H. Grunhut$^{1,4}$, M. Shultz$^{1,4}$, 
\newauthor{J.-C. Bouret$^{5,6}$, A. Fullerton$^7$, W. Marcolino$^8$, F. Martins$^9$, Y. Naz\'e$^{10}$, A. ud Doula$^{11}$,}
\newauthor{N.R. Walborn$^{7}$, J.-F. Donati$^{12}$ and the MiMeS Collaboration}\thanks{E-mail: wade-g@rmc.ca}\\
$^1$Dept. of Physics, Royal Military College of Canada, PO Box 17000, Stn Forces, Kingston, Ontario K7K 7B4, Canada \\
$^2$Dept. of Physics and Astronomy, UCL, Gower Place, London WC1E 6BT, United Kingdom \\
$^3$Dept. of Astronomy, University of Wisconsin-Madison, 475 N. Charter Street, Madison WI 53706-1582, USA\\
$^4$Dept. of Physics, Engineering Physics and Astronomy, Queen's University, 99 University Avenue, Kingston, Ontario K7L 3N6, Canada\\
$^5$LAM-UMR 6110, CNRS \& Universit\' de Provence, rue Fr\'ed\'eric Joliot-Curie, F-13388 Marseille Cedex 13, France\\
$^6$NASA/GSFC, Code 665, Greenbelt, MD 20771, USA\\
$^7$Space Telescope Science Institute, 3700 San Martin Drive, Baltimore, MD 21218, USA\\
$^8$Observatório do Valongo, Universidade Federal do Rio de Janeiro, Rio de Janeiro, Brazil\\
$^9$LUPM-UMR5299, CNRS \& Universit\'e Montpellier II, Place Eug\`ene Bataillon, F-34095, Montpellier, France\\
$^{10}$FNRS-Institut d'Astrophysique et de G\'eophysique, Universit\'e de Li\`ege, Belgium\\
$^{11}$Penn State Worthington Scranton, 120 Ridge View Drive, Dunmore, PA, USA 18512\\
$^{12}$Observatoire Midi-Pyr\'en\'ées, 14 avenue Édouard Belin, F-31400, Toulouse, France }

\begin{document}

\date{Accepted . Received , in original form }

\pagerange{\pageref{firstpage}--\pageref{lastpage}} \pubyear{2002}

\maketitle

\label{firstpage}

\begin{abstract}
This paper reports high-precision Stokes $V$ spectra of HD 191612 acquired using the ESPaDOnS spectropolarimeter at the Canada-France-Hawaii Telescope, in the context
of the Magnetism in Massive stars (MiMeS) Project. Using measurements of the equivalent width of the H$\alpha$ line and radial velocities of various metallic lines, we have updated both the spectroscopic and orbital ephemerides of this star. We confirm the presence of a strong magnetic field in the photosphere of HD 191612, and detect its variability. We establish that the longitudinal field varies in a manner consistent with the spectroscopic period of 537.6 d, in an approximately sinusoidal fashion.  The phases of minimum and maximum longitudinal field are respectively coincident with the phases of maximum and minimum H$\alpha$ equivalent width and $H_{\rm p}$ magnitude. This demonstrates a firm connection between the magnetic field and the processes responsible for the line and continuum variability. Interpreting the variation of the longitudinal magnetic field within the context of the dipole oblique rotator model, and adopting an inclination $i=30\degr$ obtained assuming alignment of the orbital and rotational angular momenta, we obtain a best-fit surface magnetic field model with obliquity $\beta=67\pm 5\degr$ and polar strength $B_{\rm d}=2450\pm 400$~G . The inferred magnetic field strength implies an equatorial wind magnetic confinement parameter $\eta_*\simeq 50$, supporting a picture in which the H$\alpha$ emission and photometric variability have their origin in an oblique, rigidly rotating magnetospheric structure resulting from a magnetically channeled wind. This interpretation is supported by our successful Monte Carlo radiative transfer modeling of the photometric variation, which assumes the enhanced plasma densities in the magnetic equatorial plane above the star implied by such a picture, according to a geometry that is consistent with that derived from the magnetic field. Predictions of the continuum linear polarisation resulting from Thompson scattering from the magnetospheric material indicate that the Stokes $Q$ and $U$ variations are highly sensitive to the magnetospheric geometry, and that expected amplitudes are in the range of current instrumentation.\end{abstract}


%
%

\section{Introduction}

The classification Of?p was first introduced by Walborn (1972) to describe spectra of early O-type stars exhibiting the presence of C~{\sc iii} $\lambda 4650$ emission with a strength comparable to the neighbouring N~{\sc iii} lines. Well-studied Of?p stars are now known to exhibit recurrent, and apparently periodic, spectral variations (in Balmer, He~{\sc i}, C~{\sc iii} and Si~{\sc iii} lines), narrow P Cygni or emission components in the Balmer lines and He~{\sc i} lines, and UV wind lines weaker than those of typical Of supergiants (see Naz\'e et al. 2010 and references therein). 

The subject of this paper, HD 191612, is perhaps the best-studied of the 5 known Galactic Of?p stars (Walborn et al. 2010). Walborn ({1973}) was the first to note the distinctive peculiarities in its spectrum. Large, recurrent spectral variations were discovered by Walborn et al. (2003). Subsequently, Walborn et al. (2004) found that these variations were strictly correlated with the low amplitude $\sim540$~d variations of the Hipparcos lightcurve (Koen \& Eyer 2002), suggesting an underlying "clock" that was proposed by Walborn et al. (2004) to be a binary orbit.  However, the detection of a magnetic field in HD 191612 by Donati et al. (2006, $B_\ell=-220\pm 38$~G) led to speculation that the observed variability was in fact rotational modulation, driven by the magnetic field and its interaction with the stellar wind. { In this scenario, the magnetic, photometric and spectral variations may be interpretable within the context of the Oblique Rotator Model (ORM; Stibbs 1950). In the ORM, a large-scale magnetic field (typically a dipole) is "frozen" into the stellar plasma, and tilted relative to the stellar rotation axis. As the star rotates, observable quantities (e.g. the line-of-sight component of the magnetic field, stellar brightness, emission lines) are modulated according to the rotational period.}

Systematic spectral and photometric studies in the optical and X-ray domains have been carried out by Howarth et al. (2007) and Naz\'e et al. (2007). Naz\'e et al. (2007) analysed phase-resolved X-ray and optical observations of HD 191612. They find that the star is overluminous in X-rays relative to the canonical $\log L_{\rm x}-\log L_{\rm bol}$ relation (by about a factor of 5), that it appears brighter in X-rays when the optical emission lines are strongest, that the X-ray lines are quite broad and that the spectrum is dominated by a cool, thermal component. Based on those and new observations, Naz\'e et al. (2010) confirmed that the X-ray emission is modulated according to the $\sim 540$~d period. These authors also performed a temporal variance spectrum (TVS) analysis on their optical spectra, confirming that the most significant spectral variability was associated with Balmer and He~{\sc i} lines (Walborn et al. 2003). They noted that the TVS profiles of Balmer and He~{\sc i} lines are roughly Gaussian, of constant phase, and centred at the rest wavelength. The line profiles consist of an apparently constant absorption component with superimposed variable emission extending from approximately -200 to 200 \kms\ . In contrast to the Balmer and He lines, absorption lines from He~{\sc ii} and absorption/emission lines of metals exhibit double-peaked TVS profiles which are more typical of binary signatures. 


From an extensive spectroscopic dataset spanning 17~y, Howarth et al. (2007) demonstrate unambiguously that the equivalent widths (EWs) of all variable spectral lines can be reasonably phased according to a unique period of $537.6\pm 0.4$~d. In particular, the striking H$\alpha$ EW variation exhibits strict periodicity and is characterised by a single, relatively sharp emission maximum and a broader, flatter minimum. The H$\alpha$ extrema are separated in phase by exactly one-half cycle. When the Hipparcos photometry (ESA, 1997) of HD 191612 are phased with the H$\alpha$ ephemeris, they also exhibit a coherent sinusoidal variation, with extrema located at the phases of the H$\alpha$ extrema. Howarth et al. also established that HD 191612 is an eccentric double-lined spectroscopic binary (SB2) with an orbital period of $1542\pm 14$~d and a mass ratio of $0.483\pm 0.044$. Those authors examined the relationship between the 537.6~d and 1542~d periods, concluding that they are independent, and that the binary orbit has no important role in the spectral variability in the optical. This is also the case in the X-ray domain (Naz\'e et al. 2010). In Table 1 we summarise the physical and wind properties of HD 191612 as reported by Howarth et al. (2007) and as derived in this study.  

\begin{table}
\centering
\caption{Summary of stellar, wind, magnetic and magnetospheric properties of HD~191612.}
\begin{tabular}{l|ll}
\hline
Spectral type & O6f?p - O8fp     & Walborn et al. (2010)         \\
$T_{\rm eff}$ (K) & 35 000 $\pm$ 1000 & Howarth et al. (2007)\\
log $g$ (cgs) & 3.5 $\pm$ 0.1  & Howarth et al. (2007)   \\
R$_{\star}$ (R$_\odot$) & 14.5 & Howarth et al. (2007)\\
$v\sin i$ (km\,s$^{-1}$) & $\ltsim 60$ & Howarth et al. (2007)  \\
$\log (L_\star/L_\odot)$ & $5.4$  & Howarth et al. (2007)\\
$M_{\star}$ ($M_{\odot}$) & $\sim 30$ & Howarth et al. (2007)\\
\hline
{$\log \dot{M}\sqrt{f}$} (M$_{\odot}$\,yr$^{-1}$) & $-5.8$  & Howarth et al. (2007)\\
$v_{\infty}$ (km\,s$^{-1}$) & 2700  & Howarth et al. (2007)\\
\hline
$B_{\rm d}$ (G) & $2450\pm 400$ & This paper\\
$\beta$ ($\degr$) & $67\pm 5$ & This paper\\
\hline
$\eta_*$ & 50 & This paper\\
$W$ & $2\times 10^{-3}$ & This paper\\
$\tau_{\rm spin}$ & 0.33 Myr & This paper\\
\hline\hline
\end{tabular}
\label{params}
\end{table}


In this paper we report systematic monitoring of the magnetic field and spectrum of HD 191612 within the context of the Magnetism in Massive Stars (MiMeS) Project (e.g. Wade et al. 2011). In Sect. 3 we describe the observations obtained, and update the ephemerides corresponding to the 537.6~d period and the 1542~d period. In Sect. 4 we analyse the magnetic data using the Least Squares Deconvolution method, and diagnose the magnetic field. In Sect. 5 we examine the period content of the (variable) longitudinal magnetic field, and demonstrate that the data can be modeled as a periodic, sinusoidal signal, with a period equal to the 537.6~d period inferred from spectroscopy. In Sects. 6 and 7 we constrain the strength and geometry of the magnetic field dipole component, and find that the geometry constraint derived from the magnetic data is in excellent agreement with the geometry obtained by Howarth et al. (2007) from modeling of the H$\alpha$ variation. In Sect. 8 we evaluate the capability of the light variation to constrain the stellar geometry, and ultimately adopt a reference geometry in Sect. 9. Finally, in Sect. 10 we discuss the implications of these results for our understanding of the behaviour of HD 191612, and for this general class of magnetic O-type stars.

\section{Observations}

High resolution spectropolarimetric (Stokes $I$ and $V$) observations of HD 191612 were collected with ESPaDOnS at the Canada-France-Hawaii Telescope. Altogether, 24 observations were obtained over a time span of 1590 days. The majority of the observations were obtained between July 2008 and October 2010; one sequence was also obtained in June 2006. Typically, two observations were acquired per night, during 2 nights each observing semester. The observing cadence was derived under the tentative assumption that the 537.6~d spectroscopic and photometric period represented the rotational period of the star, and would therefore represent the period of variation of the magnetic field. Each spectropolarimetric sequence consisted of four individual subexposures taken in different polarimeter configurations.

From each set of four subexposures we derived Stokes $I$ and Stokes $V$ spectra following the double-ratio procedure described by Donati et al. (1997), ensuring in particular that all spurious signatures are removed at first order. Null polarization spectra (labeled $N$) were calculated by combining the four subexposures in such a way that polarization cancels out, allowing us to verify that no spurious signals are present in the data (see Donati et al. 1997 for more details on the definition of $N$). All frames were processed using the CFHT's Upena pipeline, feeding the automated reduction package Libre ESpRIT (Donati et al. 1997). The peak signal-to-noise ratios (SNRs) per 2.6 \kms\  velocity bin in the reduced spectra range from 400-750, with the variation due principally to weather conditions. 

The log of CFHT observations is presented in Table 2.

\begin{table*}
\caption{Log of ESPaDOnS observations of HD 191612. All exposures were 4800 seconds duration except those corresponding to odometer \#s 851550-851562, which were 2240 s. Listed are the unique CFHT odometer number of the first spectrum of each 4-subexposure sequence, the heliocentric Julian date of the midpoint of the observation, the peak signal-to-noise ratio per 2.6~\kms\  velocity bin, the epoch and phase of the observation (according to the spectroscopic ephemeris of Howarth et al. 2007, Eq. 1), the evaluation of the detection level of a Stokes $V$ Zeeman signature (DD=definite detection, MD=marginal detection, ND=no detection), and the derived longitudinal field and longitudinal field detection significance $z$ from both $V$ and $N$. In no case is any marginal or definite detection obtained in the $N$ profiles. Also included are the H$\alpha$ equivalent widths and radial velocities measured from the { averaged} spectra { acquired on each observing night}, and described in Sect. 3. The radial velocities and EW corresponding to odometer \#s 851550-851562 were published previously by Howarth et al. (2007), and we include those values here for completeness.}
\begin{center}
\begin{tabular}{ccccccrrrrrrcc}\hline\hline
                                       &                             &       &             &            &                &   \multicolumn{2}{c}{$V$} & \multicolumn{2}{c}{$N$} & \multicolumn{3}{c}{Radial velocity} & H$\alpha$ EW\\
Odometer \#               & HJD                     &SNR& Epoch& Phase & Detect & $B_\ell\pm \sigma_B$&$z$ & $B_\ell\pm \sigma_B$&$z$ &$\lambda$5800&$\lambda$6700&O$\;$\sc{ii} \\
                                      &                             & pix$^{\rm -1}$      &             &            &                &   (G) & & (G) & & \multicolumn{3}{c}{(\kms\ )}& (\AA)\\
\hline   
851550 & 2453895.98750 & 505 & 0.894 & 0.894 & DD &$-607\pm  167 $& -3.6 &   $268 \pm 167   $ &  1.6 &  +2.0 & -17.5 & -23.4  & -3.17 \\
851554 & 2453896.01572 & 499 & 0.894 & 0.894 & DD &$-724\pm  168 $& -4.3 &   $158 \pm 168   $ &  0.9 &   \\
851558 & 2453896.04396 & 528 & 0.894 & 0.894 & DD &$-551\pm  163 $& -3.4 &   $-94 \pm 165   $ & -1.9 &   \\
851562 & 2453896.07218 & 527 & 0.894 & 0.894 & MD &$-465\pm  155 $& -3.0 &   $-123 \pm 157   $ & -1.7 &   \\
1012622 & 2454674.00146 & 694& 2.341 & 0.341& ND & $ -48\pm   83$&-0.6  &   $  14\pm   83 $ & 0.2&  -7.7 & -35.3 & +6.5  &+1.32 \\
1012626 & 2454674.05947 & 698& 2.341 & 0.341& ND & $  80\pm   82$& 1.0   &  $  80\pm   82 $ & 1.0                         &      \\
1019811 & 2454697.97739 & 599& 2.386 & 0.386& ND & $  15\pm   91$& 0.2  &   $  35\pm   92 $ & 0.4&  -7.1 & -36.2 & +6.1  &+1.46  \\
1020480 & 2454701.93208 & 550& 2.393 & 0.393& ND & $  53\pm  100$& 0.5   &  $   5\pm  100 $ & 0.1 &  -6.6 & -36.7 & +7.2  &+1.51 \\
1020484 & 2454701.98999 & 543& 2.393 & 0.393& ND & $  96\pm  105$& 0.9   &  $ -54\pm  105 $ &-0.5                         &      \\
1076976 & 2454960.06239 & 505& 2.873 & 0.873& DD & $-337\pm  165$&-2.0  &   $-179\pm  164$  &-1.1&  -9.8 & -32.4 & +2.2  &-3.03   \\
1095327 & 2455017.04698 & 704& 2.979 & 0.979& DD & $-355\pm  116$&-3.1  &   $-300\pm  117$  &-2.6 &  -9.7 & -29.5 & +0.3  &-4.21  \\
1095331 & 2455017.10517 & 745& 2.979 & 0.979& DD & $-569\pm  107$&-5.3  &   $  60\pm  107 $ & 0.6                        &        \\
1116039 & 2455082.83005 & 688& 3.101& 0.101 & DD & $-414\pm  110$&-3.8   &  $-102\pm  110$  &-0.9&  -7.3 & -27.9 & -5.8  &-3.28  \\
1116043 & 2455082.88801 & 687& 3.102 & 0.102& DD & $-526\pm  113$&-4.7  &   $ -23\pm  113  $&-0.2                        &       \\
1120384 & 2455101.85585 & 686& 3.137 & 0.137& DD & $-518\pm  107$&-4.8  &   $-150\pm  107$  &-1.4&  -6.0 & -26.2 & -6.3  &-2.54   \\
1120388 & 2455101.91356 & 658& 3.137 & 0.137& DD & $-336\pm  111$&-3.0  &   $ -71\pm  111  $&-0.6                        &      \\
1204158 & 2455350.94273 & 635& 3.600 & 0.600& ND & $-151\pm   86$&-1.8  &   $  46\pm   86  $& 0.5& +15.3 & -12.9 &-38.5  &+1.43  \\
1204377 & 2455352.10185 & 690& 3.602 & 0.602& ND & $  -1\pm   78$& 0.0  &   $  40\pm   78 $ & 0.5& +15.5 & -12.4 &-39.9  &       \\
1217470 & 2455401.96936 & 435& 3.695& 0.695 & ND & $-176\pm   89$&-2.0   &  $ -22\pm   89$ & -0.3 & +9.65 & -16.6 &-30.6  &+0.82  \\
1217474 & 2455402.02757 & 402& 3.695 & 0.695& ND & $-125\pm   97$&-1.3  &   $ -58\pm   97$ & -0.6                         &       \\
1218699 & 2455407.81551 & 705& 3.706 & 0.706& DD & $-498\pm  115$&-4.3  &   $-310\pm  114 $& -2.7 & +8.95 & -17.2 &-30.6 &+0.56   \\
1218703 & 2455407.87376 & 662& 3.706 & 0.706& MD & $-208\pm  123$&-1.7  &   $  54\pm  124 $&  0.4                        &        \\
1251479 & 2455486.76659 & 659& 3.853 & 0.853& DD & $-145\pm  150$&-1.0  &   $  56\pm  150$ &  0.4  & -0.15 & -21.4 &-19.6  &-2.50   \\
1251483 & 2455486.82470 & 629& 3.853 & 0.853& DD & $  23\pm  158$& 0.1  &   $-161\pm  157$  & -1.0                        &  \\\hline\end{tabular}
\end{center}
\end{table*}

\section{Ephemerides}

\subsection{H$\alpha$}
As described in detail by Howarth et al. (2007), line profiles of HD 191612 are variable in both EW and radial velocity (see in particular their Figs. 1, 3 and 6).  We have measured the equivalent width of the H$\alpha$ profile from the mean spectra acquired on each night.  For completeness, we combined these data with the results of Howarth et al. (2007) to give the updated ephemeris summarised in Table 3. It is sensibly indistinguishable from the equivalent Howarth et al. result ($P=537.6$~d), which we adopt throughout the remainder of this paper. In particular, there is no phase drift with respect to the previous result. 

\begin{table}
\caption{Parameters of the updated H$\alpha$ ephemeris obtained from combining the new spectroscopic data with those reported by Howarth et al. (2007).}
\begin{center}
\begin{tabular}{lrcll}
\hline
$W_0$ & 2.50         &$\pm$& 0.17&\AA\\
$A$ & 6.75                && 0.15&\AA\\
$P_\alpha$ & 537.2               && 0.3 &d\\
$t_0$ & JD$\;$2,453,415.1 && 0.5 &\\
$\sigma_\phi$ & 0.177     && 0.004&\\
$\phi_0$ &   0.338        && 0.005\\
\hline
\hline
\end{tabular}
\end{center}
\end{table}

Differences between these values and those reported in Table 1 of
Howarth et al. (2007) are negligible, and do not justify formally updating the
ephemeris at this time.

\subsection{Orbit}

The new CFHT spectra are of very high quality, and now span a periastron
passage in the long-period, low-amplitude spectroscopic orbit described by Howarth et al. (2007).
Velocity measurements have been made from the mean CFHT spectra acquired on each night, and additionally from a spectrum obtained in service
mode on JD 2455264.76, close to the expected time of periastron passage, with the FIES spectrograph
on the 2.56m Nordic Optical Telescope ($388-598$~nm, $R=46,000$).


The orbital solution uses the C~{\sc iv} doublet, $\lambda\lambda$5800, to establish the primary's $\gamma$ velocity; $\lambda\lambda$5800 plus a complex of 8 emission lines (N~{\sc ii}, Si~{\sc iv}, and unidentified) around $\lambda\lambda$6700 to determine all other orbital parameters for the primary; and a set of 6 O~{\sc ii} lines ($\lambda\lambda$4300--4700, with separations constrained to laboratory values) to determine $K_2$ (and $\gamma_2$; see Howarth et al. 2007 for further details). { The typical external errors on the new velocity measurements are $\sim$1~\kms\ for C~{\sc iv} and $\sim$2--3~\kms\ for $\lambda\lambda$6700 and O~{\sc ii}.}

The revised orbit, which affords a modest improvement over
the previous solution because of the availability of superior data at
critical phases, is summarised in Table 4 and illustrated in Fig. 1.

\begin{table}
\caption{Updated orbital solution.}
\begin{center}
\begin{tabular}{lrcll}
\hline
$\gamma$ (\civ) & $-4.73$   &$\pm$&  0.32  & \kms\ \\
$K_1$         &  13.18         &&  0.72  & \kms\ \\
$e$           &  0.492         &&  0.030 \\
$\omega$      &  350.65        &&  4.1   & $^\circ$\\
$P_{\rm orb}$ &  1548.3        &&  7.4   &d \\
$T_0$         &  JD 2453744    &&  11    \\
$f(m)$        &  0.243         &&  0.042 &{\msun} \\
$a_1\sin{i}$  &  351           &&  20    &\rsun\ \\
$\Delta\gamma$ ($\lambda\lambda$6700)& 24.84 &&0.49&\kms\ \\
rms residual & \multicolumn{2}{l}{(weight 1, \civ)} &2.4&\kms\ \\
rms residual & \multicolumn{2}{l}{($\lambda\lambda$6700)} &3.3&\kms\ \\
\\
$\gamma$ (\oii) & +8.92 && 0.56& \kms\ \\
$K_2$      & 26.6&&0.8 & \kms\ \\
$f(m)$        &  1.99          &&  0.19  &{\msun} \\
$a_2\sin{i}$  &  708           &&  22    &\rsun\ \\
$q = M_2/M_1$&        0.496 &&0.031\\
rms residual & \multicolumn{2}{l}{(\oii)} & 3.5&\kms\ \\
\hline\hline
\end{tabular}
\end{center}
\end{table}

\begin{figure*}
\center{\includegraphics[scale=0.6,angle=-90]{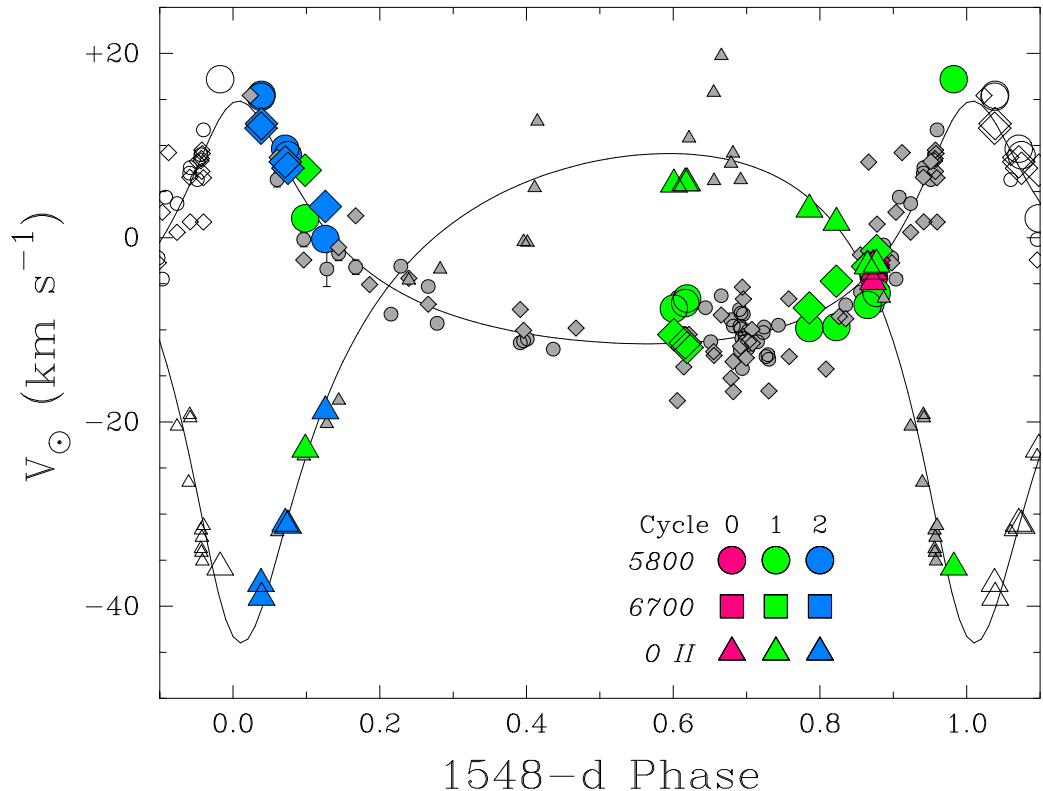}} 
\caption[]{Updated orbital solution.  New results are shown in larger
  symbols, coded according to orbital cycle/line (colour/shape).  The 
spring 2010 periastron passage is quite well covered.
$\lambda$6700
  and \oii\ results have been adjusted to the \civ\ $\gamma$ velocity.}
\end{figure*}


The new CFHT RV measurements are reported in Table 2. The FIES spectrum
yields the following velocities: +17.2 \kms\  for $\lambda 5800$, -37.9 \kms\ for O~{\sc ii}.

\section{Magnetic diagnosis}

An examination of the reduced spectra reveals weak Zeeman signatures visible in profiles of strong absorption lines during phases of peak H$\alpha$ emission. However, the SNR of these signatures is not sufficient for detailed modeling. Instead, we will combine the signal from a handful of lines in the spectrum to improve our ability to diagnose the magnetic field.

Least-Squares Deconvolution (LSD, Donati et al. 1997) was applied to all observations. In their detection of the magnetic field of HD 191612, Donati et al. (2006) developed and applied a { manually-constructed} line mask containing 12 lines. { In developing this mask, they excluded lines in emission, strong Balmer lines, and other features showing P Cygni profiles.} To begin, we used this line list to extract, for all collected spectra, mean circular polarization (LSD Stokes $V$), mean polarization check (LSD $N$) and mean unpolarized (LSD Stokes $I$) profiles. All LSD profiles were produced on a spectral grid with a velocity bin of 36 \kms, providing reasonable sampling of the observed mean profile and maximising the S/N per LSD pixel. The LSD Stokes $I$ profile shows an extension to high ($\sim 300$~\kms) velocities in the blue wing. This asymmetry is similar to that observed in the LSD profiles of the Of?p star HD~148937 (Wade et al. 2011) although not quite as strong. It is likely reflective of the inclusion of asymmetric He~{\sc i} lines in the mask. An additional dip in the blue wing is attributed to the diffuse interstellar band (DIB) located at $\sim 579.8$~nm, which was not included in the line mask but which is blended with C~{\sc iv} $\lambda$5801, which is included. 

Using the $\chi^2$ signal detection criteria described by Donati et al. (1997), we evaluated the significance of the signal in both Stokes $V$ and in $N$. In no case is any signal detected in $N$, while signal in $V$ is detected marginally (false alarm probability ${\rm fap}<10^{-3}$) or definitely (${\rm fap}<10^{-5}$) in 15 of our 24 observations. Finally, from each set of LSD profiles we measured the mean longitudinal magnetic field in both $V$ and $N$ using the first moment method (Rees \& Semel 1979) as expressed by Eq. (1) of Wade et al. (2000), integrating from -280 to +180 \kms\ . The longitudinal field measured from Stokes $V$ is detected significantly (i.e. $|z|=|B_\ell|/\sigma\ge 3$) in 11 of our observations. In no case is the longitudinal field significantly detected in $N$.

The results of the analysis described above are summarised in Table 2.

To test the sensitivity of these results to the detailed mask composition, we extracted LSD profiles for two additional lines masks. We began using a generic line mask based on a 40000~K {\sc extract stellar} request from the VALD database, using a line depth threshold of 0.1. We sought to construct a mask that yields a mean LSD Stokes $I$ line profile that exhibited as little variability as possible, and that was as symmetric as possible, while still maximizing Stokes $V$ signal. At each step of the mask development, we visualised the agreement of the LSD model (i.e. the convolutions of the Stokes $I$ and $V$ LSD profiles with the line mask) with the reduced spectrum, and evaluated the symmetry and variability of the LSD profile. We quickly realised that many lines, even if in absorption during some phases of the 537.6 d cycle, needed to be discarded from the mask due to emission at other cycle phases. After removal of the most significantly affected lines, we were left with a mask containing 26 lines, for which we adjusted the mask depths to best match the mean observed line depths. While none of the lines remaining in the mask show obvious emission, many are asymmetric and vary during the course of the 537.6~d cycle. This mask is dominated by lines of neutral He, but also contains lines of ionised He, C, N O and Si. LSD profiles extracted using this mask - which contains more than twice the number of lines in the Donati et al. mask - yields LSD profiles nearly identical to those obtained from the Donati et al. mask. This is likely a consequence of the inclusion of many of the same strong He lines in both masks. Measurements of the longitudinal fields extracted from these LSD profiles are in almost perfect agreement with those obtained above. 

As a second test, we extracted LSD profiles using a more restricted mask containing only a half-dozen metallic lines (lines of  C, O and Si). This mask yields profiles whose shapes differ substantially from those obtained from the first two masks: they are narrower, and they are nearly symmetric. This is expected due to the exclusion of the much broader, more asymmetric He~{\sc i} lines. Notwithstanding the difference in shape, however, this mask yields longitudinal fields that are in good agreement (although slightly noisier due to the weaker and more limited sample of lines) with those obtained from the masks containing He lines. Similar agreement is obtained whether we employ an integration range adapted to the (narrower) width of the metallic lines, or if we use the (larger) width adapted to the masks containing He lines.

Based on this analysis, we conclude that while the detailed shapes of the LSD profiles depend on the composition of the mask, the inferred longitudinal field of HD 191612 is relatively insensitive to the mask composition. We therefore based our following analysis on the LSD profiles extracted using the 12 line mask of Donati et al. (2006).

As mentioned above, we typically obtained two observations of HD 191612 during each observing night (or, in one case, within several nights). We confirmed that no significant variability had occurred between any of these near-simultaneous observations. This is consistent with the known long variability period of this star. To improve the precision of our measurements, in our following analysis of the magnetic variability and geometry of HD 191612 we have coadded and remeasured LSD profiles obtained close in time (i.e. on the same night or, in the case of observations 1204158 and 1204177, within 3 nights). Ultimately, this has yielded 12 independent measurements of the magnetic field. These measurements, which will be used for the remainder of the analysis, are reported in Table 5. Examples of the coadded LSD profiles are shown in Fig. 2.

In our following analysis, we also include the observation of Donati et al. (2006) of  $-220\pm 38$~G obtained on HJD 2453546.55279.    

\begin{figure*}
\centering
\includegraphics[width=8.5cm]{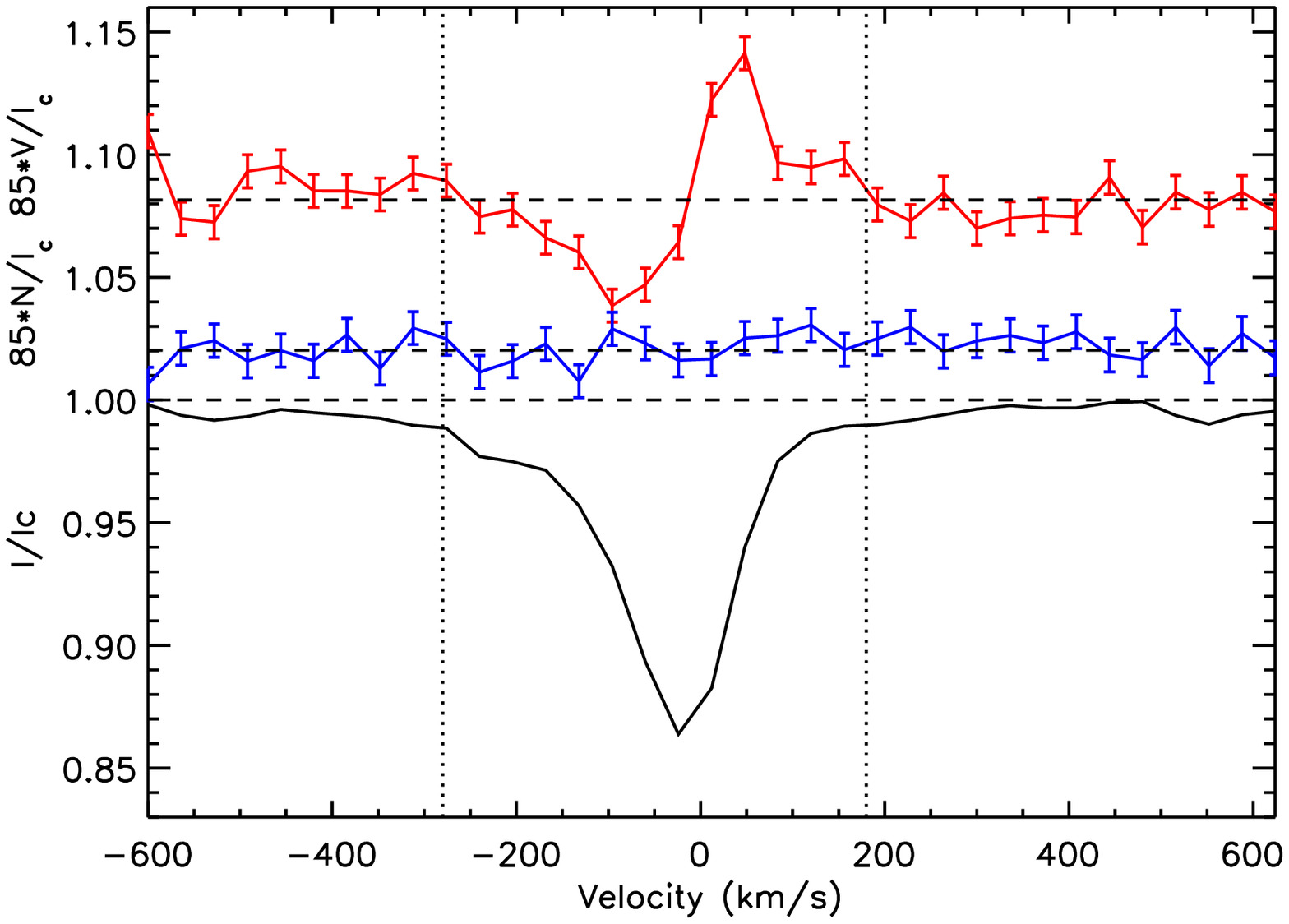}\includegraphics[width=8.5cm]{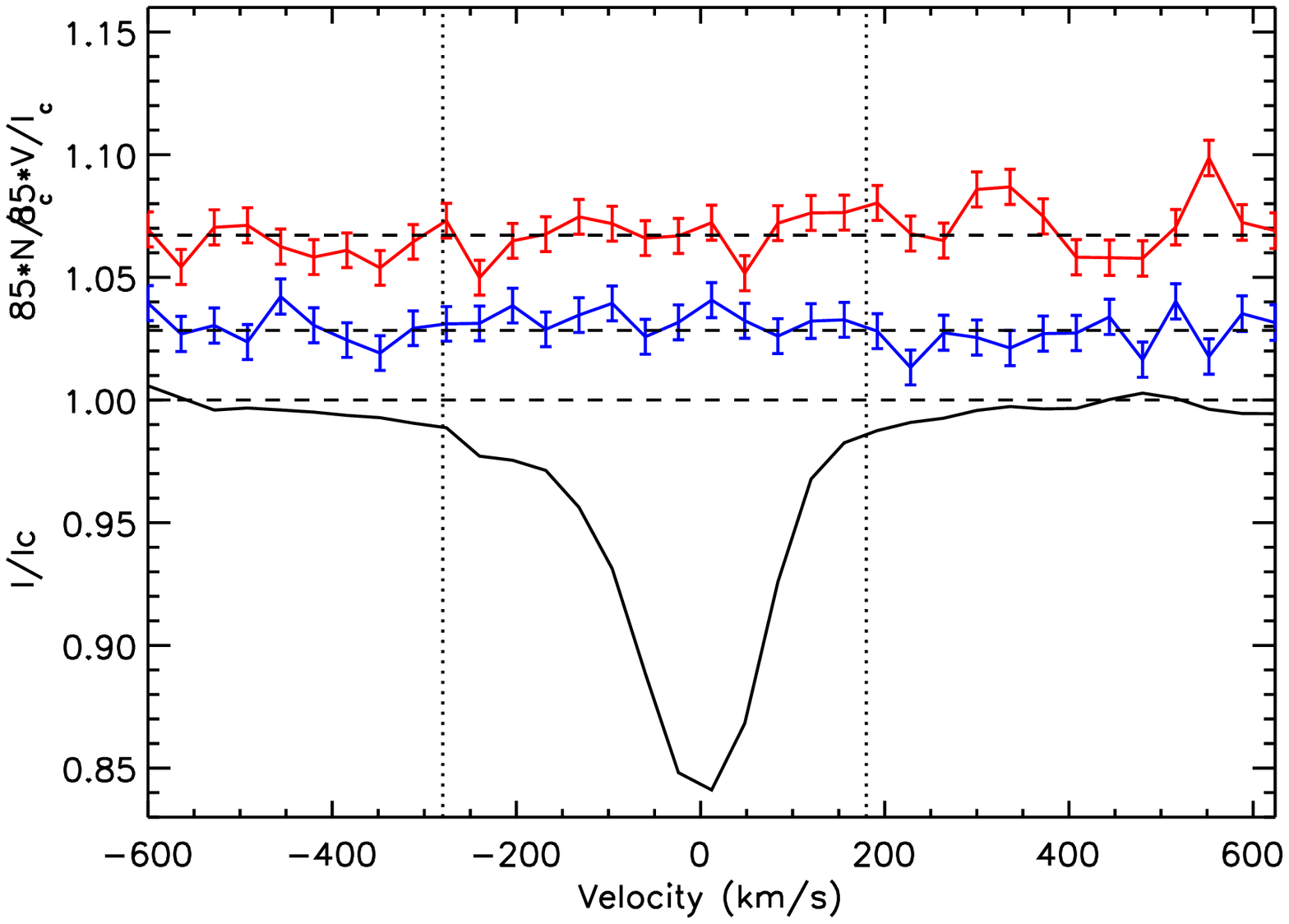}
\caption{LSD profiles obtained using the Donati et al. mask. {\em Left:}\ Stokes $V$, $N$ and $I$ LSD profiles at phase 0.979 ($B_\ell=-514 \pm  81$).  {\em Right:}\ Stokes $V$, $N$ and $I$ LSD profiles at phase 0.601 ($B_\ell=-65 \pm 63$). The profiles have been scaled according to the SNR-weighted mean wavelength, land\'e factor and line depth of the mask. Note the significant variation of the strength of the Stokes $I$ profile.}
\label{SNRs}
\end{figure*}

\section{Magnetic period}

The longitudinal magnetic field measurements of HD 191612 are strongly inconsistent with the null field hypothesis (reduced $\chi^2$ of 19.4). This confirms the magnetic field detection reported by Donati et al. (2006). Nor are they consistent with a nonzero but nonvariable field (reduced $\chi^2$ of 8.6). The longitudinal magnetic field of HD 191612 is therefore detected to vary on the timescales of our observations, with a maximum measured value of $67\pm 73$~G and a minimum value of $-514\pm 81$~G.

Including the measurement of Donati et al. (2006), the longitudinal field measurements span a total time of 1940 days. The Lomb-Scargle periodogram of the magnetic measurements between 1 and 1000 days shows a single minimum with reduced $\chi^2$ below 2.5. The minimum is double-peaked, with dominant peak (with reduced $\chi^2$ of 0.5) at 539 d and a secondary peak (with reduced $\chi^2$ of 1.5) at 623 d, both $\pm 20$~d approximately. The 539~d period is consistent with the well-determined spectroscopic period of $537.6\pm 0.4$~d, while the 623~d period may be a sidelobe. Although the magnetic data set only spans 3.6 cycles of this length, the lack of other significant power in the periodogram supports a basic tenet of the oblique rotator model (ORM): that rotational modulation determines the variability of magnetic and other observables, and that their periods should therefore generally be the same. Indeed, when phased with the 537.6 day spectroscopic period (Howarth et al. 2007) the magnetic data display a smooth, approximately sinusoidal variation, with a very low reduced $\chi^2$ of 0.5 (which may suggest that our error bars are slightly overestimated). We note that the data do not phase satisfactorily according to the SB2 orbital period of 1542~d (or other periods in this vicinity), yielding a best reduced $\chi^2\sim 6.5$).

We therefore conclude that the variable magnetic field of HD 191612 is best interpreted within the context of the ORM. In this case, the common magnetic/spectroscopic/photometric period of 537.6 d is the rotational period of the star. The magnetic observations therefore confirm the (until now) tentatively assumed ORM framework (e.g. Donati et al. 2006, Howarth et al. 2007) for interpretation of the observed variability, and the extreme slow rotation of HD 191612. Moreover, extending these conclusions to the cases of other Of?p stars in which magnetic fields have been detected (HD 108 [Martins et al. 2010], HD 148937 [Wade et al. 2011]) supports the view that all magnetic Of?p stars host strong, organised magnetic fields, that they may well be oblique rotators, and that their diverse spectroscopic periods (from 7~d for HD 148937 to perhaps 55~y for HD 108) likely represent their real rotational periods.

\section{Phase variation of the longitudinal field}




To model the magnetic field of HD 191612 within the framework of the oblique rotator model, we begin by phasing all of the data according to the H$\alpha$ ephemeris of Howarth et al. (2007):

\begin{equation}
{\rm JD =} 2453415.2(5) \pm 537.6(4)\cdot {\rm E}.
\end{equation}

The phased longitudinal field measurements are illustrated in the upper frame of Fig. 3. 

To verify the phasing of the magnetic data relative to the spectroscopic ephemeris, we have also measured the equivalent width of H$\alpha$ from each of the co-added spectra, using the same methods described by Howarth et al. (2007). The new H$\alpha$ measurements phase perfectly with the extensive dataset published by Howarth et al., indicating that no significant phase drift has occurred since the acquisition of those observations. The phased H$\alpha$ data (both new and old) are illustrated in the middle frame of Fig. 3.

What is immediately apparent in Fig. 3 is that the extrema of the H$\alpha$ equivalent width variation, as well as the Hipparcos photometric variation (lower frame), occur simultaneously with the extrema of the magnetic variation.  This further strengthens the view implicit to the oblique rotator model that there exists a causal relationship between the magnetic field variation, and the spectral and photometric variations.

We note that Hubrig et al. (2010) report a single SOFIN observation of the longitudinal field of HD 191612, $B_\ell=+450\pm 153$~G, acquired on September 11, 2008, and corresponding to rotational phase 0.43. This strong positive longitudinal field does not agree { very} well with our observed field variation, which corresponds to (an essentially) consistently negative longitudinal field. 


{ Hubrig et al. also report the presence of very strong (up to 15\%) Stokes $V$ signatures in the Ca~{\sc ii} H and Na~{\sc i} D lines that have the same (positive) sign as the stellar longitudinal field they infer.  However, these lines are interstellar features in the spectrum of HD 191612, and consequently a linkage with the stellar field is unlikely.  We observe no similar Stokes signatures in our ESPaDOnS spectra of these lines.  The strong interstellar signatures might arise from the non-standard normalization of the Stokes $V$ spectrum, which Hubrig et al. indicate to be $V/I$ rather than the more usual $V/I_c$); see their Fig. 1. }


\begin{figure*}
\centering
\includegraphics[width=12cm,angle=-90]{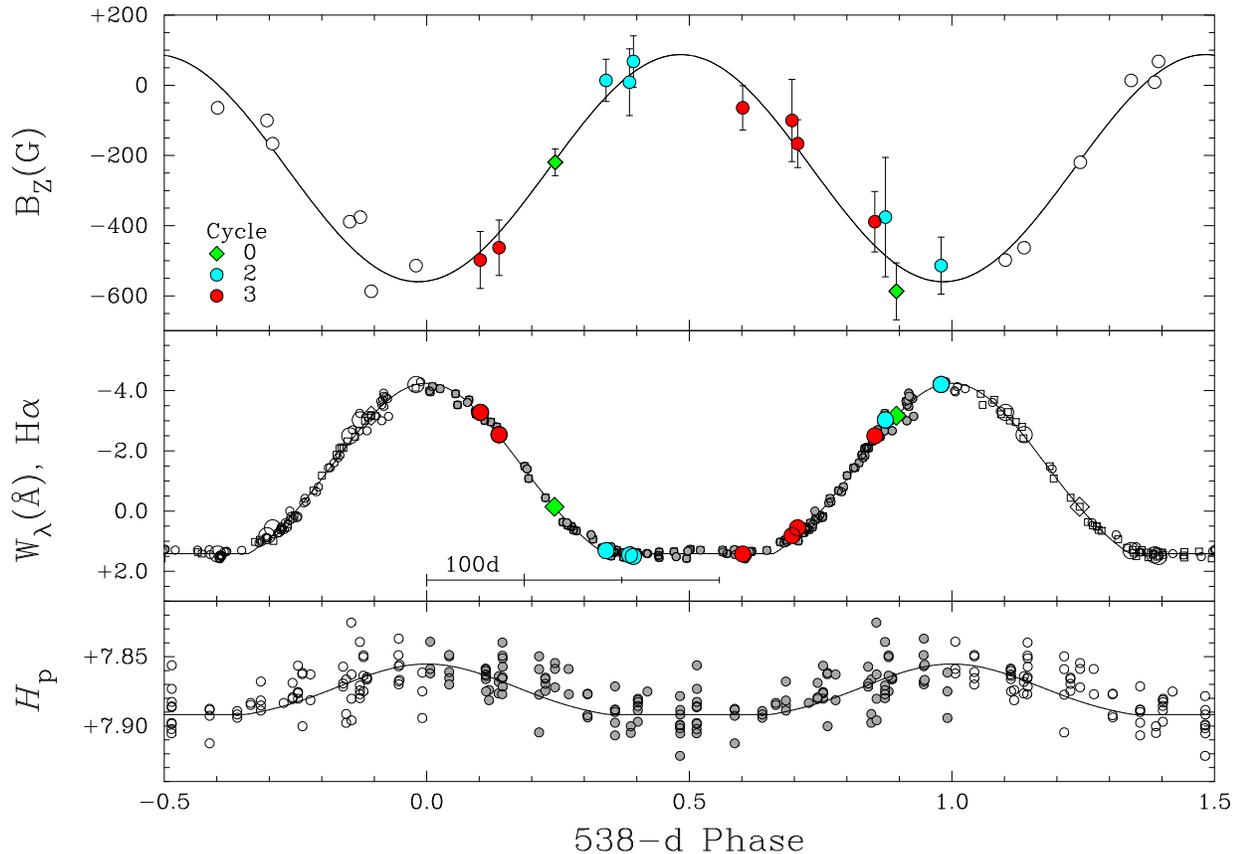}
\caption{Longitudinal magnetic field, H$\alpha$ EW and Hipparcos magnitude, all phased according to the 537.6d period. The measurement of Donati et al. (2006) appears at phase 0.23. Solid lines are least-squares fits to the data.}
\label{SNRs}
\end{figure*}



\section{Stellar geometry implied by the magnetic variation}

The least-squares sinusoidal fit $B_\ell(\phi)=B_0+B_1\cos 2\pi(\phi+\phi_1)$ to the phased magnetic data yields coefficients $B_0=-236\pm 10$~G, $B_1=324\pm 20$~G, and $\phi_1=0.48\pm 0.04$ corresponding to extrema of $B_\ell^{\rm max}=+88\pm 22$~G, $B_\ell^{\rm min}=-560\pm 22$~G. { The reduced $\chi^2$ of this fit is 0.5, suggesting an overestimation of the longitudinal field error bars by a factor of $\sim 1.4$, assuming a target reduced $\chi^2$ of unity. This may be a consequence of the relatively small number of lines used in the LSD line mask.}

We model the magnetic field of HD 191612 assuming an oblique rotating dipole, characterised by four parameters: the phase of closest approach of the magnetic pole ($\phi_1-1=-0.02$, obtained from the least-squares fit), the stellar rotation axis inclination ($i$) to the observer's line-of-sight, the magnetic axis obliquity ($\beta$) and the dipole polar strength $B_{\rm d}$. The longitudinal field variation provides constraint on two of these unknowns, typically $\beta$ and $B_{\rm d}$. Frequently, the inclination $i$ is inferred using the measured projected rotational velocity and period, and the inferred stellar radius. However, in the case of HD 191612, only an upper limit on $v\sin i$ is known ($\leq 60$~\kms), and this value is clearly highly in excess of the true $v\sin i$ if $P_{\rm rot}=537.6$~d. We therefore cannot infer $i$ using the usual methods. 

To begin, assuming the inclination angle to be unconstrained, we have modeled the longitudinal field variation for $i$ ranging from $20-80\degr$, obtaining best fit values $13\degr\leq\beta\leq 75\degr$ and $2000$~G$\leq|B_{\rm d}|\leq 5000$~G. For inclination angles approaching $0/90\degr$, the obliquity approaches zero and the dipole strength diverges. On the other hand, models with $i=0\degr$ and $i=90\degr$ are not able to reproduce the longitudinal field variation. The general family of solutions is characterised by $i+\beta=95\pm 10\degr$.

Howarth et al. (2007) attempted to reproduce the form of the H$\alpha$ EW variation using two simple models: first, a centred, tilted, geometrically thin but optically thick H$\alpha$-emitting disc, and secondly, a single surface spot. Both are equally capable of reproducing the H$\alpha$ variation, and both models yield the basic geometrical constraint $i+\alpha\simeq 100\degr$, where $\alpha$ is the obliquity of the disc or the colatitude of the spot\footnote{$105\degr$ was stated by Howarth et al., but a revisit of the modeling by IH yields the slightly lower $100\degr$.}. This constraint is consistent with that derived from the longitudinal field variation, substituting $\beta$ for $\alpha$, i.e. that the $\alpha$ angle employed by Howarth et al. (2007) is in fact equal to the magnetic obliquity. In the particular case of the thin disc model, this implies that the disc lies in the plane of the magnetic equator, as would be expected in the framework of a magnetically-channeled wind (e.g. Babel \& Montmerle 1997, ud Doula \& Owocki 2002). 

\begin{figure*}
\centering
\includegraphics[width=16cm]{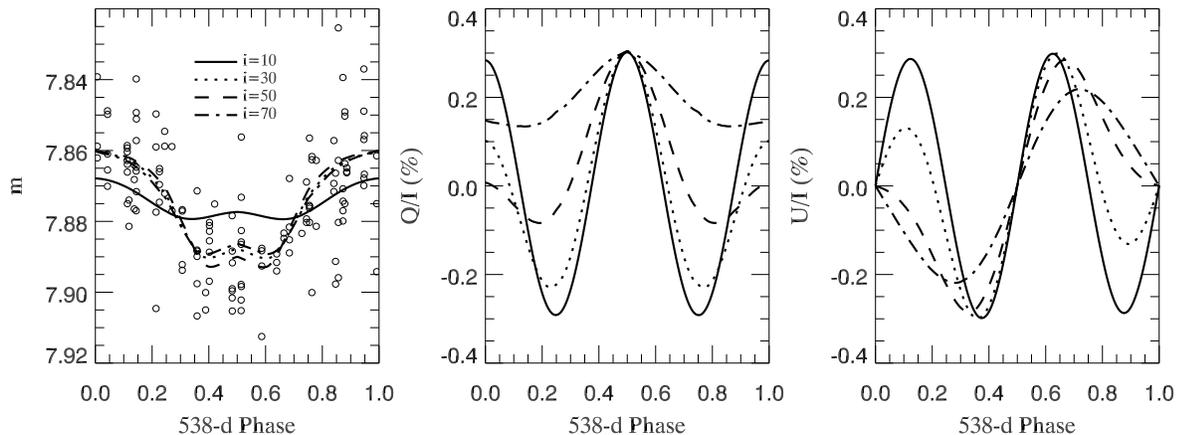}
\caption{Photometric (left) and linear polarisation (centre and right) phase variations of HD 191612 computed for various rotational axis inclinations. The open circles in the left-hand panel represent the Hipparcos photometry. The Hipparcos photometry of HD 191612 constrains $i>30\degr$.}
\label{SNRs}
\end{figure*}

\begin{table}
\caption{Averaged data used for modeling.}
\begin{center}
\begin{tabular}{ccccrrc}\hline\hline
Mean HJD           & Epoch              & Phase    & Detect&          $B_\ell \pm \sigma_B$ & $z$ \\
                               &                           &                 &            &          (G)            & \\
\hline
2453896.030 & 0.894  & 0.894    & DD &   $-587\pm  81$ & -7.3 \\
2454674.031 & 2.341&  0.341 &MD& $  13 \pm  60$    &  0.2    \\
2454697.977 & 2.386&  0.386 &ND& $   8 \pm  95$    &  0.1    \\
2454701.961 & 2.393&  0.393 &ND& $  67 \pm  73$    &  0.9   \\
2454960.062 & 2.873&  0.873 &DD& $-376 \pm 170$    & -2.2      \\
2455017.076 & 2.979&  0.979 &DD& $-514 \pm  81$    & -6.4    \\
2455082.859 & 3.102&  0.102 &DD& $-498 \pm  81$    & -6.2    \\
2455101.885 & 3.137&  0.137 &DD& $-463 \pm  79$    & -5.9     \\
2455351.522 & 3.601&  0.601 &ND& $ -65 \pm  63$    & -1.0    \\
2455401.999 & 3.695&  0.695 &ND& $-101 \pm 117$    & -0.9    \\
2455407.845 & 3.706&  0.706 &DD& $-167 \pm  68$    & -2.5     \\
2455486.796 & 3.853&  0.853 &DD& $-389 \pm  86$    & -4.5   \\ 
\hline\hline
\end{tabular}
\end{center}
\end{table}

\section{Monte Carlo radiative transfer simulation of the light variation}

To leverage the diagnostic potential offered by the \emph{Hipparcos} light curve (Fig. 3), we use a Monte-Carlo radiative transfer (RT) code developed by RHDT for simulating light scattering in circumstellar envelopes. The RT code follows much the same procedure as other recent codes (see, e.g., Wood \& Reynolds 1999).  Photon packets are launched from a central star and allowed to propagate through an arbitrary distribution of circumstellar matter (described by a Cartesian density grid), until they are scattered by free electrons. We neglect the possibility of packet absorption, since at temperatures similar to the $T_{\rm eff}$ of HD~191612 the optical (Paschen continuum) bound-free cross section is negligible compared to the Thomson cross section. Upon scattering, a ray is peeled off from the packet toward a virtual observer, who records the packet's Stokes parameters appropriately attenuated by any intervening material (see Yusuf-Zadeh, Morris \& White 1984, for a discussion of this peel-off technique). A new propagation direction is then chosen based on the dipole phase function (Chandrasekhar 1960), and the packet's Stokes parameters are updated to reflect the linear polarization introduced by the scattering process. The propagation is then resumed until, after possible further scatterings, the packet eventually escapes from the system or is reabsorbed by the star.

The initial emission direction of photon packets, relative to the local stellar-surface normal, is determined randomly in accordance with the limb-darkening law tabulated by Chandrasekhar (1960) for plane-parallel electron scattering atmospheres. The corresponding initial Stokes parameters are determined from the same tabulation. To improve the efficiency of the code, all packets are forced to undergo at least one scattering, using the formalism described by Witt (1977).

In applying the RT code to HD~191612, we establish the circumstellar density distribution from a 2-D (axisymmetric) MHD simulation of the star's magnetically channeled wind using the ZEUS 3-D code (see ud-Doula \& Owocki 2002 and Owocki \& ud-Doula 2004, for a general overview of our simulation approach) using the full energy equation as in Gagn\'e et al. (2005), and the parameters summarised in Table 1. To smooth out temporal variations in the density, we take an average over 1,000\,Ms of simulation time (corresponding to $\sim 30$ wind flow times), beginning at 300\,ks after the initial state. Because the star rotates so slowly, the dynamical effects of the rotation are negligible, and we can use the same density distribution (albeit suitably realigned to the magnetic axis) for any choice of the obliquity $\beta$. Electron number densities are calculated from the mass density under the assumption of solar composition and complete ionization.

Fig.~4 shows the results from RT simulations for inclinations $i=10\degr,30\degr,50\degr$ and $70\degr$, and corresponding obliquities $\beta =85\degr,65\degr,45\degr$ and $25\degr$ (these are the values mandated by the $i+\beta=95\degr$ relation; see Sec. 7). The left-hand panel shows the light curves over one rotation cycle, for the four choices of inclination. All models correctly reproduce the distinctive shape of the \emph{Hipparcos} light curve. However, only the $i \geq 30\degr$ models are able also to match the observed $\sim 4\,{\rm mmag}$ amplitude, allowing us
to rule out the $i=10\degr$ case. (We stress that the amplitudes of our model light curves are \emph{not} free parameters, but rather \emph{predictions} of the MHD and RT simulations for the adopted stellar properties).

From the light curves alone, it is not possible to distinguish between the $i=30\degr, 50\degr$ and $70\degr$ models. However, the centre and right-hand panels of Fig.~4, which plot the models' Stokes $Q$ and $U$ parameters over one rotation cycle, clearly reveal that the linear polarization variations of the star are sensitive to inclination. Moreover, the presence and variation of the linear polarisation would implicitly confirm the presence of a flattened, disc-like structure. This finding establishes a strong case for obtaining fresh phase-resolved linear continuum polarization observations of HD~191612. Such observations - which require a precision of order 0.01\% - are within range of current instrumentation (e.g. Carciofi et al. 2007).

\section{Tentative reference geometry}

Although the inclination $i$ is not strongly constrained by the measurements in hand, the data provide a relatively precise diagnosis of the magnetic parameters $\beta$ and $B_{\rm d}$ for a given value of $i$. Given our desire to present a reference model for the magnetic field of HD 191612, we proceed as follows. Hypothesising that the component masses in the binary system are more or less normal for their spectral types (O8fp for the primary, B1V for the secondary), this suggests that $\sin(i_{\rm orb})$ is very likely within the range 0.4-0.6 (Fig. 9 of Howarth et al. 2007). Speculating that the orbital and rotational angular momenta are aligned, we tentatively adopt $i=i_{\rm orb}\simeq 30\degr$ as a first guess at the rotation axis inclination. We note that this inclination is consistent with the observed Hipparcos photometric variation according to the modeling described in the previous section. 

Modeling the longitudinal field variation using a numerical implementation of the dipole oblique rotator model, for $i=30\degr$, we obtain $B_{\rm d}=2450\pm 400$~G and $\beta=67\pm 5\degr$. For illustration, the reduced $\chi^2$ landscape for $\beta$ vs. $B_{\rm d}$ corresponding to $i=30\degr$ is presented in Fig. 5. 

The results of our magnetic field modeling are in remarkably good agreement with the strength and geometry proposed by Donati et al. (2006) based on their single measurement of the longitudinal field, and guidance provided by the H$\alpha$ equivalent width variation. In particular, they proposed that the geometry of the magnetic field is constrained according to $i+\beta\simeq 90\degr$ (essentially the constraint we derive here from a complete modeling of the phase variation of the longitudinal field). Their inferred dipole strength of 1.5~kG derived for $i=\beta=45\degr$ is very similar to that which we derive for this same geometry, although somewhat weaker than that derived for our $i=30\degr$ reference geometry.



\section{Discussion}

\begin{figure}
\centering
\includegraphics[width=8cm]{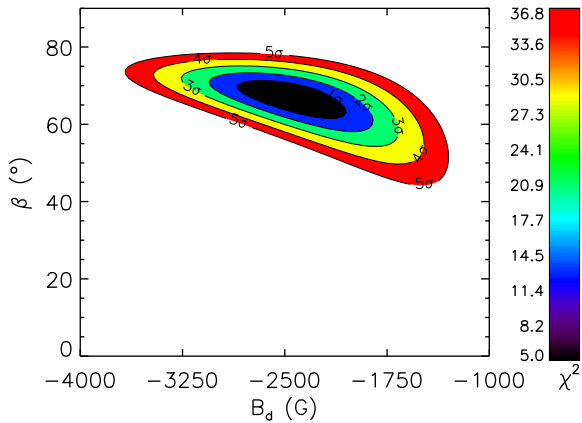}
\caption{Reduced $\chi^2$ landscape of $\beta$ vs $B_{\rm d}$ for fixed inclination $i=30\degr$. Contours correspond to 1-5$\sigma$ limits obtained from fitting the (two free parameter) oblique magnetic dipole model to the phased, averaged longitudinal field variation. We obtained $B_{\rm d}=2450\pm 400$~G and $\beta=67\pm 5\degr$, where the error bars are the formal (1$\sigma$) uncertainties.}
\label{SNRs}
\end{figure}

Our new magnetic observations of HD 191612 confirm the existence of an organised magnetic field in the photosphere of this Of?p star. We have demonstrated that the longitudinal magnetic field varies significantly, with a period that is in agreement with the spectroscopic period of 537.6 d. The observed phenomenology is consistent with an oblique magnetic rotator, for which the spectroscopic period is the stellar rotation period. 

The H$\alpha$ equivalent width and Hipparcos photometric measurements vary in phase with the longitudinal field. This strongly suggests a causal connection between their variability and the longitudinal field. In particular, H$\alpha$ { emission from, and electron scattering in, a disc of plasma located in the magnetic equator are able to qualitatively and, to some extent quantitatively,} reproduce all observables.

Modeling the longitudinal field variation and adopting a rotation axis inclination to the line of sight $i=30\degr$ (which assumes alignment of the spin and orbital angular momenta of HD 191612, and which is consistent with the lower limit on $i$ implied by our modeling of the light variations), we derive the dipole strength and obliquity $B_{\rm d}=2450\pm 400$~G and $\beta=67\pm 5\degr$. Using the stellar and wind parameters reported by Howarth et al. (2007), we compute the wind magnetic confinement parameter $\eta_*=B_{\rm eq}^2\,R^2/\dot M v_{\rm \infty}\simeq 50$ and rotation parameter $W=v_{\rm eq}/v_{\rm crit}=2\times 10^{-3}$ (e.g. ud Doula et al. 2008). This places the Alfven radius at about $R_{\rm Alf}=\eta_*^{1/4}=2.2~R_*$. \footnote{If we use the theoretical mass-loss recipe of Vink et al (2000), we obtain an unclumped mass-loss rate of $\log(\dot M)=-6.1$ (twice smaller than that found by Howarth et al. (2007), resulting in a confinement parameter of $\eta_*\sim 100$ and an Alfven radius of $R_{\rm Alf}=3.2 R_*$.} The Kepler (or corotation) radius is located much further away, at about $R_{\rm Kep}=W^{-2/3}=60~R_*$. As pointed out by ud Doula et al. (2008), for any material trapped on magnetic loops inside the Kepler radius, the outward centrifugal support is less than the inward pull of gravity; since much of this material is compressed into clumps that are too dense to be significantly line-driven, it eventually falls back to the star following complex patterns along the closed field loops. Hence, for HD 191612, all magnetically-confined plasma (i.e. all wind plasma located inside the Alfven radius) is unstable to this phenomenon. Therefore the modulation of the H$\alpha$ emission (and presumably much of the other line variability, and the photometry as well) is a consequence of a statistical overdensity of plasma near the magnetic equatorial plane inside $R_{\rm Alf}$. However, this dense plasma is not supported either by the radiative or the centrifugal forces, and thus exhibits a relatively short residence time. The "disc" of HD 191612 is therefore physically distinct from the rotationally-supported, rigidly-rotating magnetospheres of rapidly-rotating magnetic stars such as $\sigma$~Ori E (Townsend, Owocki \& Groote 2005) or HR 7355 (Oksala et al. 2010, Rivinius et al. 2010).

Naz\'e et al. (2007) highlight basic differences between the X-ray spectrum of HD 191612 and those of the known magnetic hot stars $\tau$~Sco and $\theta^1$~Ori C, both of which (like HD 191612) are characterised by $R_{\rm Alf}<<R_{\rm Kep}$. In particular, the X-ray spectrum of HD 191612 displays a relatively soft spectrum with few high-ionisation lines, broad ($\sim 2000$~\kms) lines, and lack of bremsstrahlung continuum. In contrast, $\tau$~Sco and $\theta^1$~Ori C show narrow X-ray lines (always below 1000~\kms, and typically a few hundred \kms), hard spectra with a dominant hot component and high ionisation lines.  $\theta^1$~Ori C furthermore exhibits bremsstrahlung continuum, clearly indicating that the emission measure distribution is dominated by high temperature plasma. They proposed that HD 191612 is an intermediate case between the magnetic OB stars and the "typical" O stars. The results presented here demonstrate that the magnetic and magnetospheric characterstics of HD 191612 (dipolar field topology, field strength of order kG, intermediate obliquity, $\eta_*$ and $W$ parameters) are rather similar to those of $\theta^1$~Ori C. Hence the differences in X-ray characteristics of these stars must now be understood in light of our confirmation of the magnetic oblique rotator model for HD 191612. Given the relatively high quality of the current X-ray and other observations of Of?p stars, this will likely require guidance to theory provided by a pan-spectral approach to a large and diverse sample of magnetic OB stars.

The inferred rotation period of HD 191612 - approximately 1.5 years - is extremely slow. Computing the magnetic braking spindown time according to Eq. (25) of ud Doula et al. (2009), using the parameters in Table 1, we obtain 0.33 Myr. If the age of HD 191612 is 3-4 Myr as suggested by the analysis of Donati et al. (2006), we conclude (as did Donati et al. 2006 and ud Doula et al. 2009) that magnetic braking alone could be responsible for increasing the rotational period of HD 191612 from that of a typical ZAMS early O-type star to its current value. We note however that this model of rotational spindown does not take into account the (strongly) changing stellar moment of inertia, mass loss characteristics and magnetic characteristics during the evolution of the star from the ZAMS. In particular, if the surface dipole magnetic field of HD~191612 is currently of order 2 kG, under the assumption of magnetic flux conservation it may have been larger than 30~kG at the ZAMS (or possibly even larger than that, given the results of Landstreet et al. 2008). We conclude that more sophisticated models taking into account the time-dependence of the spindown must be developed to accurately compare observed magnetic stars with spindown predictions.

The identification of this period with the period of stellar rotation assumes the validity of the ORM. Such a long rotational period implies an equatorial rotational velocity of order 1 km/s, which would not be detectable in the photospheric lines of this star. Similarly, the $\sim 55$~y period inferred for HD~108 (Martins et al. 2010), if it is in fact the rotational period, also implies a negligible $v\sin i$. The remarkable similarity in shape of some of the least variable, presumably most "photospheric" lines (those least modified by the wind or magnetosphere, e.g. O~{\sc iii} 5590) of these stars is consistent with this idea. On the other hand, those same lines in the spectrum of HD~148937, which according to the ORM should be rotating much faster (with a period of 7~d, with $v_{\rm eq}\simeq 108$~\kms; Wade et al. 2011), also show profiles that differ negligibly from HD 191612. This suggests that the profiles of those lines contain no significant information about the stellar projected rotational velocity.


Our confirmation of the oblique rotator model for HD 191612 provides a reasonable basis for proposing that other Of?p stars in which magnetic field is detected (HD 108 - Martins et al. 2010, HD 148937 - Wade et al. 2011) are also oblique rotators, with their variation periods being their rotational periods. We have proposed that the general class of Of?p stars may be the first category of O-type magnetic oblique rotators. It should be a priority to investigate the remaining Galactic Of?p stars (NGC 1624-2 and CPD$-28^{\rm o} 2561$; Walborn et al. 2010) to seek to detect magnetic fields, and investigate their rotation and wind confinement.

\section*{Acknowledgments}
GAW acknowledges support from the Natural Science and Engineering Research Council of Canada (NSERC). STScI is operated by AURA, Inc., under NASA contract NAS5-26555. { YN acknowledges support from the Fonds National de la Recherche Scientifique (Belgium), the PRODEX XMM and Integral contracts, and the `Action de Recherche Concert\'ee' (CFWB-Acad\'emie Wallonie Europe).}

\end{document}